\documentclass{PoS}

\usepackage{amsmath}
\usepackage{amssymb}
\usepackage{wrapfig}

\title{Constraints for nuclear PDFs from the LHCb D-meson data}

\ShortTitle{Constraints for nuclear PDFs from the LHCb D-meson data}

\author{Kari J. Eskola \\
        University of Jyvaskyla, Department of Physics, P.O. Box 35, FI-40014 University of Jyvaskyla, Finland \\
        Helsinki Institute of Physics, P.O. Box 64, FI-00014 University of Helsinki, Finland \\
        E-mail: \email{kari.eskola@jyu.fi}}

\author{Ilkka Helenius \\
        University of Jyvaskyla, Department of Physics, P.O. Box 35, FI-40014 University of Jyvaskyla, Finland \\
        Helsinki Institute of Physics, P.O. Box 64, FI-00014 University of Helsinki, Finland \\
        E-mail: \email{ilkka.m.helenius@jyu.fi}}

\author{Petja Paakkinen \\
        University of Jyvaskyla, Department of Physics, P.O. Box 35, FI-40014 University of Jyvaskyla, Finland \\
        Helsinki Institute of Physics, P.O. Box 64, FI-00014 University of Helsinki, Finland \\
        E-mail: \email{petja.paakkinen@jyu.fi}}

\author{\speaker{Hannu Paukkunen} \\
        University of Jyvaskyla, Department of Physics, P.O. Box 35, FI-40014 University of Jyvaskyla, Finland \\
        Helsinki Institute of Physics, P.O. Box 64, FI-00014 University of Helsinki, Finland \\
        E-mail: \email{hannu.paukkunen@jyu.fi}}

\abstract{
We quantify the impact of LHCb D-meson measurements at $\sqrt{s}=5 \, {\rm TeV}$ on the EPPS16 and nCTEQ15 nuclear PDFs. In our study, the theoretical description of D-meson production is based on the recently developed SACOT-$m_{\rm T}$ variant of the general-mass variable-flavour-number formalism, and the impact on PDFs is estimated via reweighting methods. We pay special attention on the theoretical uncertainties known to us, and are led to exclude the $p_{\rm T}<3 \, {\rm GeV}$ region from our main analysis. The LHCb data can be accommodated well within EPPS16 and nCTEQ15, and the data provide stringent constraints on the gluons in the shadowing/antishadowing regions. No evidence of non-linear effects beyond standard DGLAP evolution is found even if the full kinematic region down to zero $p_{\rm T}$ is considered.
}

\FullConference{XXVII International Workshop on Deep-Inelastic Scattering and Related Subjects - DIS2019\\
		8-12 April, 2019\\
		Torino, Italy}

\begin{document}

\section{Introduction}
\vspace{-0.2cm}

As has been shown in several recent works \cite{Zenaiev:2015rfa,Gauld:2016kpd,Kusina:2017gkz}, D-meson production at the LHC can serve as a stringent constraint for proton and nuclear parton distribution functions (PDFs). However, the theoretical treatment of D-meson production is not fully established, and the used approaches vary from the fixed flavour-number scheme (FFNS) \cite{Mangano:1991jk} to \textsc{fonll} \cite{Cacciari:1998it}, to FFNS combined with parton showers \cite{Frixione:2007nw}, and to matrix-element fitting \cite{Lansberg:2016deg}. In the work summarized here \cite{Eskola:2019bgf}, we have used the general-mass variable-flavour-number scheme (GM-VFNS) approach \cite{Kniehl:2004fy,Helenius:2018uul}.

\vspace{-0.2cm}
\section{D-meson production in GM-VFNS}
\vspace{-0.2cm}

The GM-VFNS ``master formula'' for D-meson production at fixed rapidity $Y$ and transverse momentum $P_{\rm T}$ is (see Refs.~\cite{Kniehl:2004fy,Helenius:2018uul} for details),
\begin{align}
\frac{d\sigma^{h_1h_2 \rightarrow D^0 + X}}{dP_{\rm T}dY} = \sum _{ijk}
\int \frac{dz}{z} dx_1 dx_2 
            & { f_i^{h_1}(x_1,\mu^2_{\rm fact}) }
            { {d\hat{\sigma}^{ij\rightarrow k}(x_1, x_2, m, \mu^2_{\rm ren}, \mu^2_{\rm fact}, \mu^2_{\rm frag})} } \\
            & { f_j^{h_2}(x_2,\mu^2_{\rm fact}) }            
            { D_{k \rightarrow {\rm D}^0}(z,\mu^2_{\rm frag}) }             
        \nonumber \,.
\end{align}
\begin{wrapfigure}{r}{0.45\textwidth}
\vspace{-0.7cm}
\hspace{-0.3cm}\includegraphics[width=1.05\linewidth]{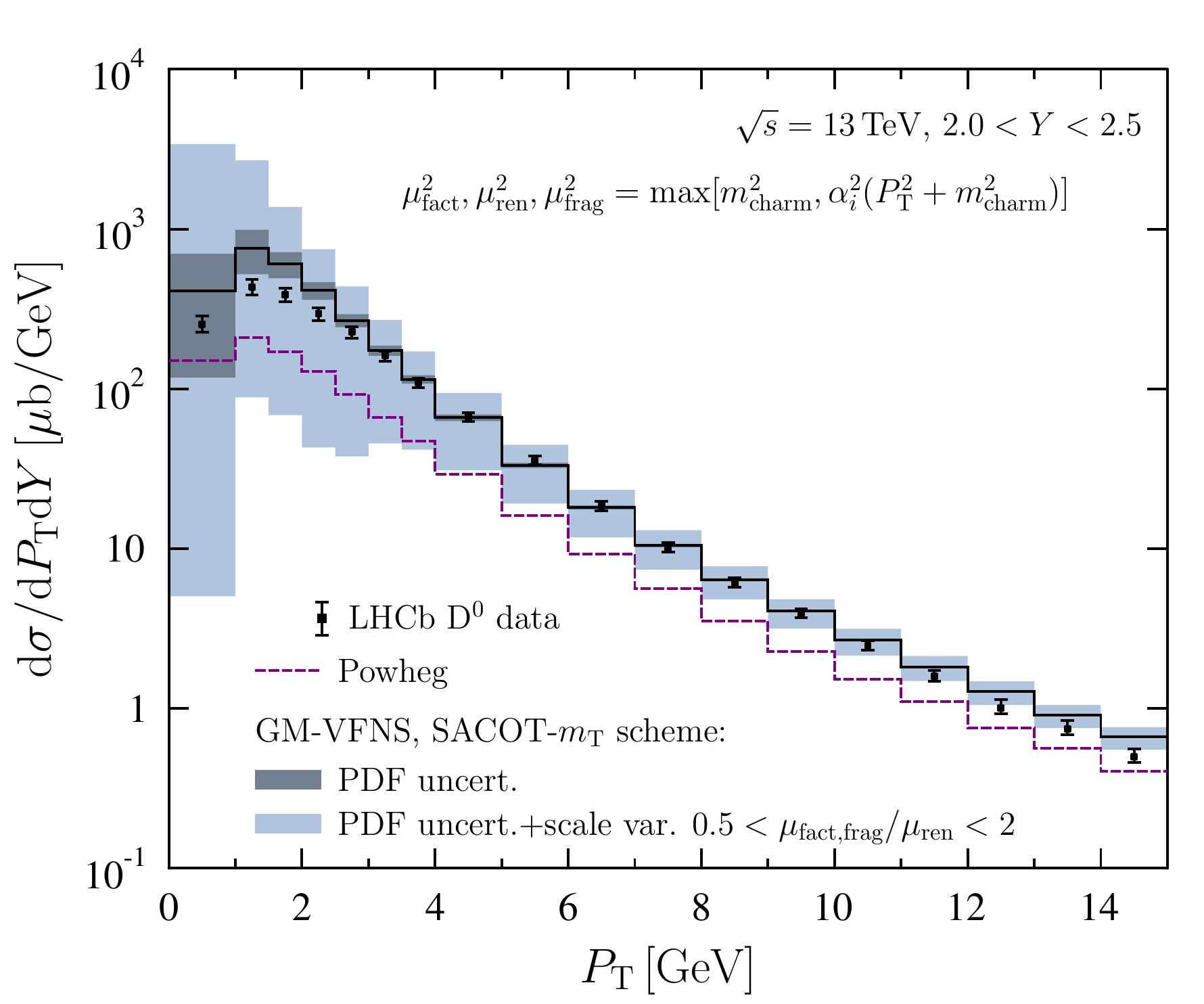}

\hspace{-0.3cm}\includegraphics[width=1.05\linewidth]{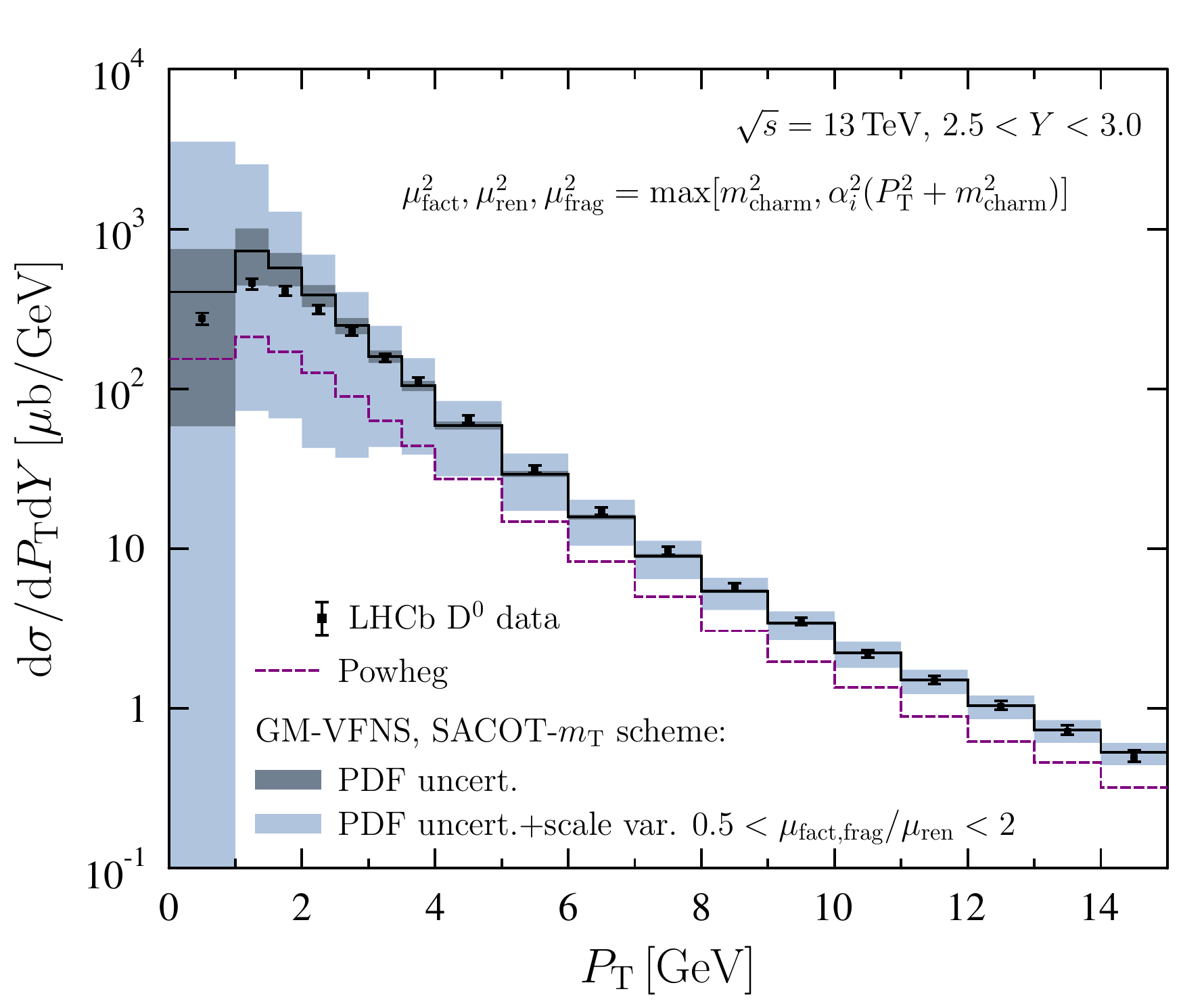}
\caption{LHCb p-p data \cite{Aaij:2015bpa} vs. SACOT-$m_{\rm T}$ calculation (solid histograms). Figures from Ref.~\cite{Helenius:2018uul}.}
\label{fig:ppspectra}
\end{wrapfigure}
Here, $f_i^{h}(x_1,\mu^2_{\rm fact})$ denote the PDFs at factorization scale $\mu^2_{\rm fact}$, $D_{k \rightarrow {\rm D}^0}(z,\mu^2_{\rm frag})$ are the parton-to-meson fragmentation functions
(FFs) at fragmentation scale $\mu^2_{\rm frag}$, and ${d\hat{\sigma}^{ij\rightarrow k}}$ are the partonic Wilson coefficients which also depend on the renormalization scale $\mu^2_{\rm ren}$ and charm-quark mass $m$. Towards small $P_{\rm T}$, the coefficient functions behave as in the FFNS, but at the high-$P_{\rm T}$ limit the zero-mass $\overline{\rm MS}$ expressions are recovered. The fragmentation variable $z$ is not unique due to the finite masses of the charm quark and the D meson. In the above formula, we have defined $z$ as the fraction of the fragmenting partons' energy carried by the D meson. The GM-VFNS formulation differs from e.g. FFNS in including all partonic subprocesses (FFNS includes only $g+g \rightarrow Q\overline{Q} + X, \, q+\overline{q} \rightarrow Q\overline{Q} + X, \, q+g \rightarrow Q\overline{Q} + X$) and in that the FFs are scale dependent. As is well known, the GM-VFNS entails a scheme dependence related to the exact way the mass dependence of the contributions initiated by heavy quarks and those in which a light quark or gluon fragments into a D meson, is treated.
In our work, we have adopted the so-called SACOT-$m_{\rm T}$ scheme \cite{Helenius:2018uul}, in which the scheme-dependent contributions retain the kinematics of the heavy-quark pair production, but the Wilson coefficients are massless. In this scheme, the heavy-quark mass regulates the cross sections at small $P_{\rm T}$ leading to a judicious description from zero to asymptotically high $P_{\rm T}$. This is demonstrated in Figure~\ref{fig:ppspectra} where the SACOT-$m_{\rm T}$ framework is compared with the LHCb p-p data \cite{Aaij:2015bpa} -- the behaviour of the data is clearly well described. 

\vspace{-0.5cm}
\section{Nuclear modifications in p-Pb collisions}
\vspace{-0.2cm}

The LHCb collaboration has measured differential nuclear-modification factors
\begin{equation}
R_{\rm pPb}^{\rm D^0} \equiv \frac{d\sigma^{\rm pPb \rightarrow D^0 + X}/dP_{\rm T}dY}{d\sigma^{\rm pp \rightarrow D^0 + X}/dP_{\rm T}dY} 
\end{equation}
\begin{figure}[b!]
\center
\vspace{-0.2cm}
\includegraphics[width=0.48\linewidth]{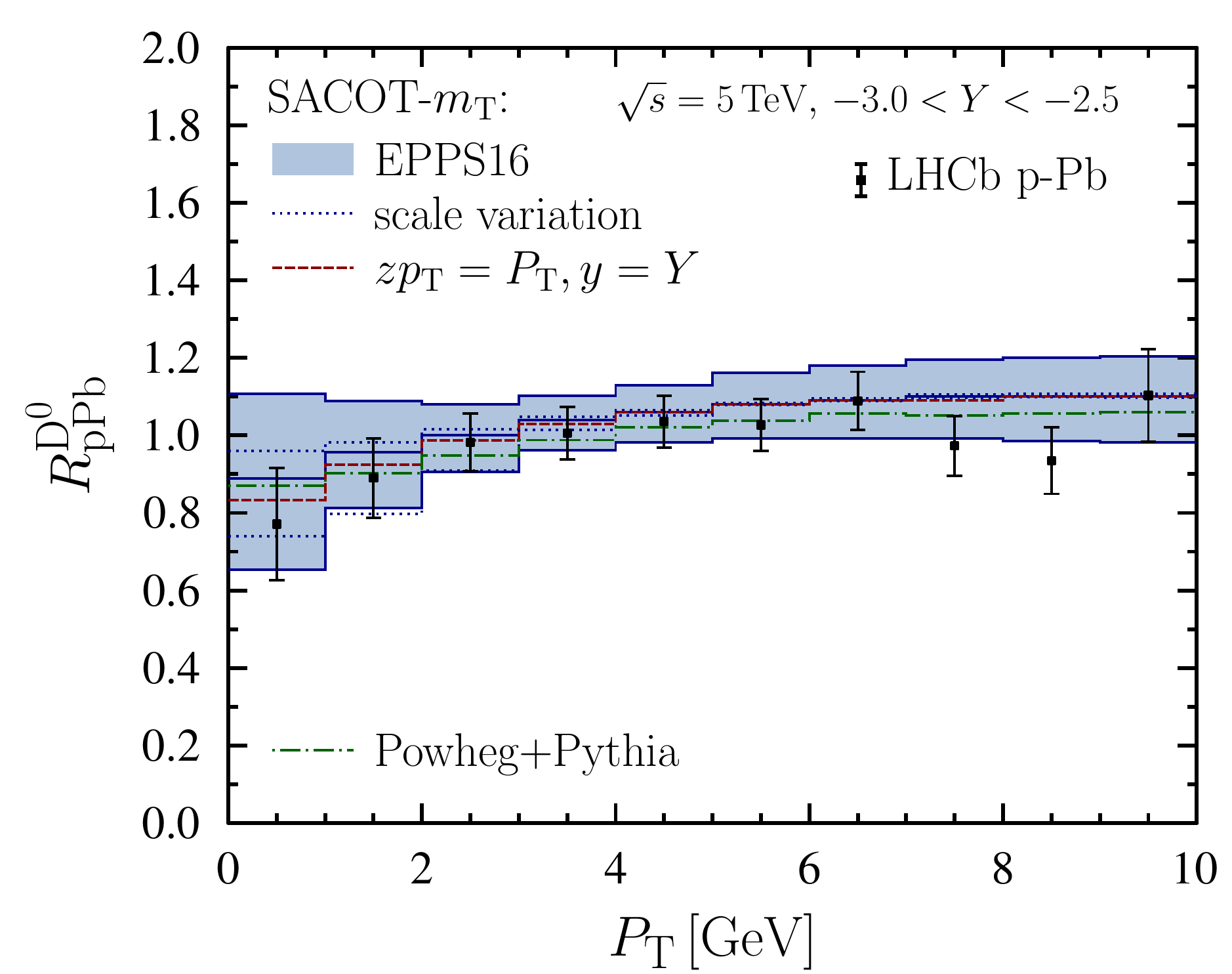}
\includegraphics[width=0.48\linewidth]{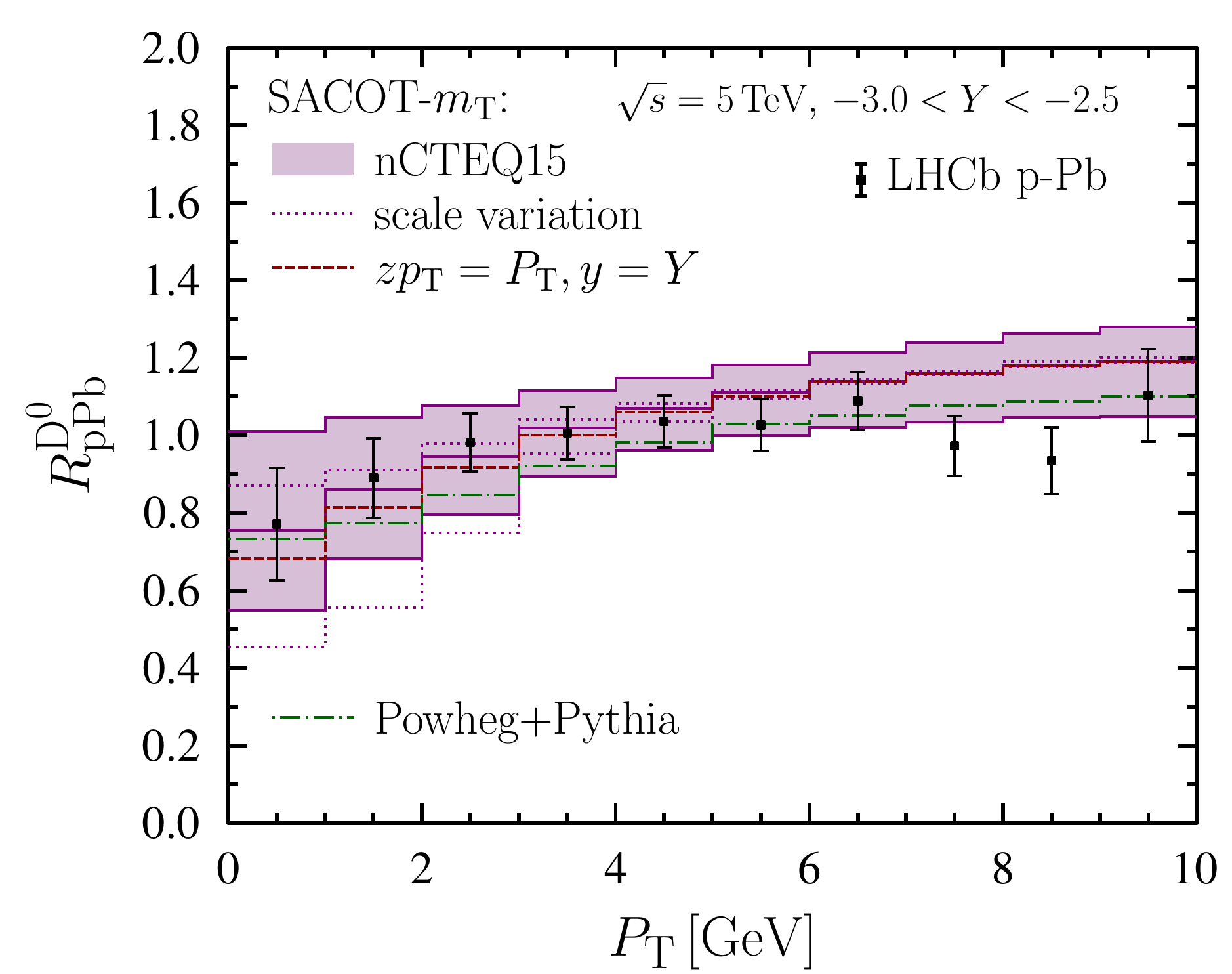} \\
\includegraphics[width=0.48\linewidth]{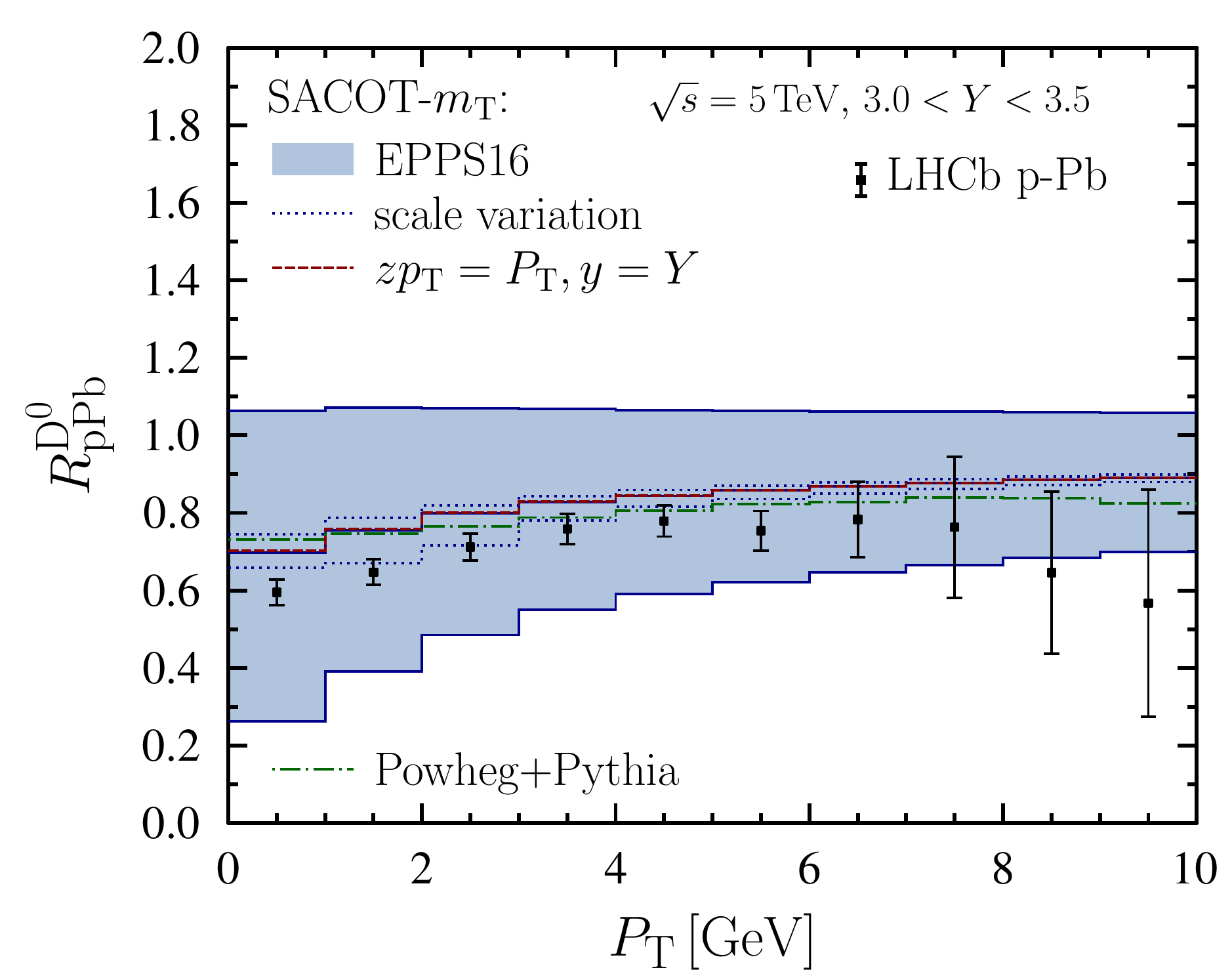}
\includegraphics[width=0.48\linewidth]{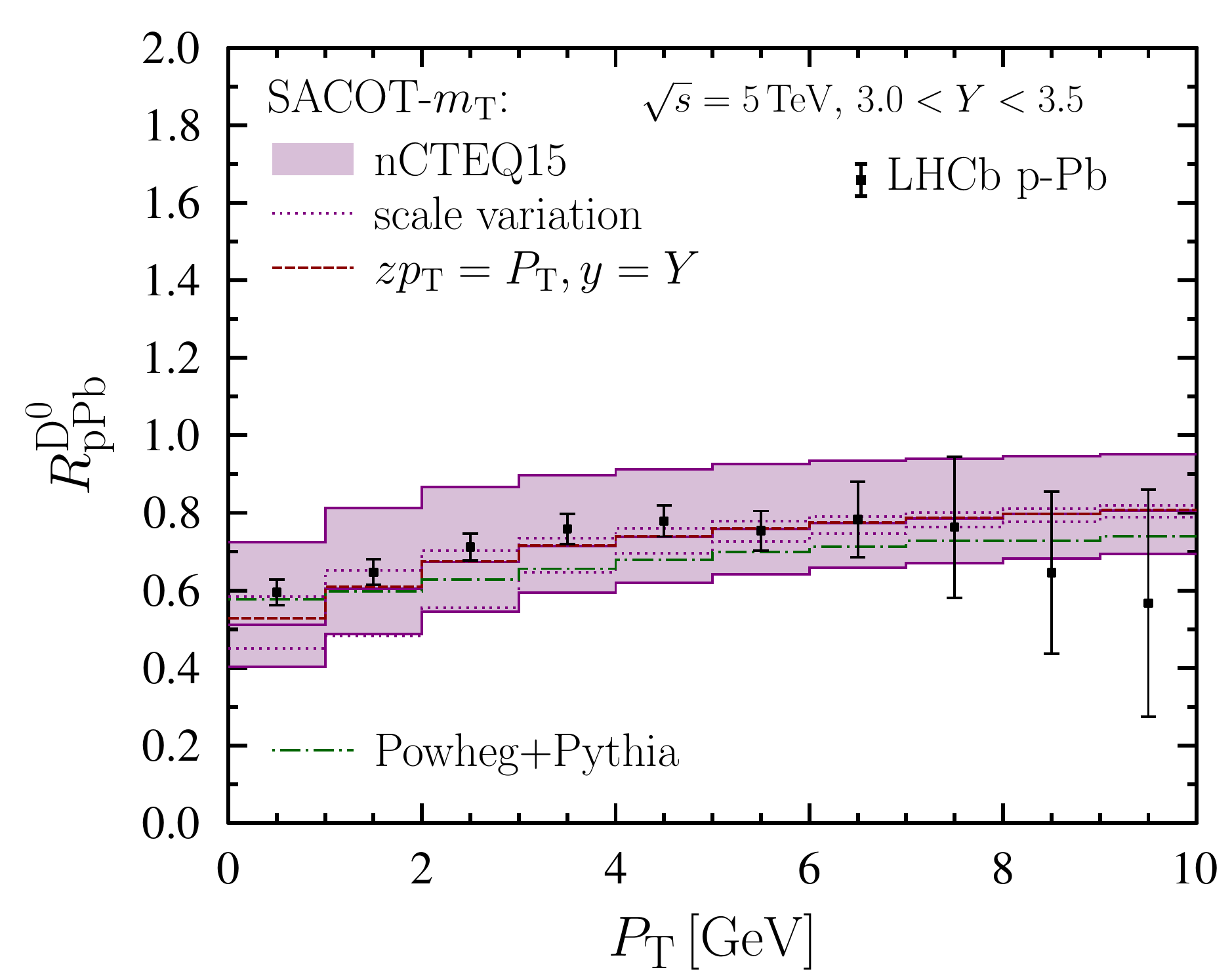}
\caption{LHCb data \cite{Aaij:2015bpa} for the nuclear modification ratio $R^{\rm D^0}_{\rm pPb}$ compared with EPPS16 (left) and nCTEQ15 (right) predictions. }
\label{fig:pPb1}
\end{figure}
at $\sqrt{s}=5 \, {\rm TeV}$ for D$^0$ mesons \cite{Aaij:2017gcy}. In Figure~\ref{fig:pPb1}, we compare the measurements at backward ($-3 < Y <-2.5$) and forward ($3 < Y < 3.5$) directions to our calculations using the EPPS16 \cite{Eskola:2016oht} and nCTEQ15 \cite{Kovarik:2015cma} nuclear PDFs. We have investigated the theoretical uncertainties in these ratios by varying the scales $\mu^2_{\rm ren}, \mu^2_{\rm fact}, \mu^2_{\rm frag}$ (dotted lines), and also tried the zero-mass prescription for the fragmentation variable $z$ (dashed lines). As we can see from Figure~\ref{fig:pPb1}, at high $P_{\rm T}$ the uncertainties of these origins get negligible, but below $P_{\rm T} \sim 3\,{\rm GeV}$ the variation begins to be significant. We have also computed predictions by using first \textsc{Powheg} \cite{Frixione:2007nw} to generate $c\overline{c}$ events and then hadronizing them with \textsc{Pythia} \cite{Sjostrand:2014zea}. These predictions (dashed-dotted lines) tend to be below the GM-VFNS predictions. The underlying reason is that by generating only events where the heavy-quark pair is explicitly produced at the hard scattering, the Powheg+Pythia approach samples the PDFs at overly small $x$ leading to a systematic bias in the predictions \cite{Helenius:2018uul}.

\vspace{-0.2cm}
\section{Reweighting analysis}
\vspace{-0.2cm}

\begin{figure}[b!]
\center
\vspace{-0.2cm}
\includegraphics[width=0.48\linewidth]{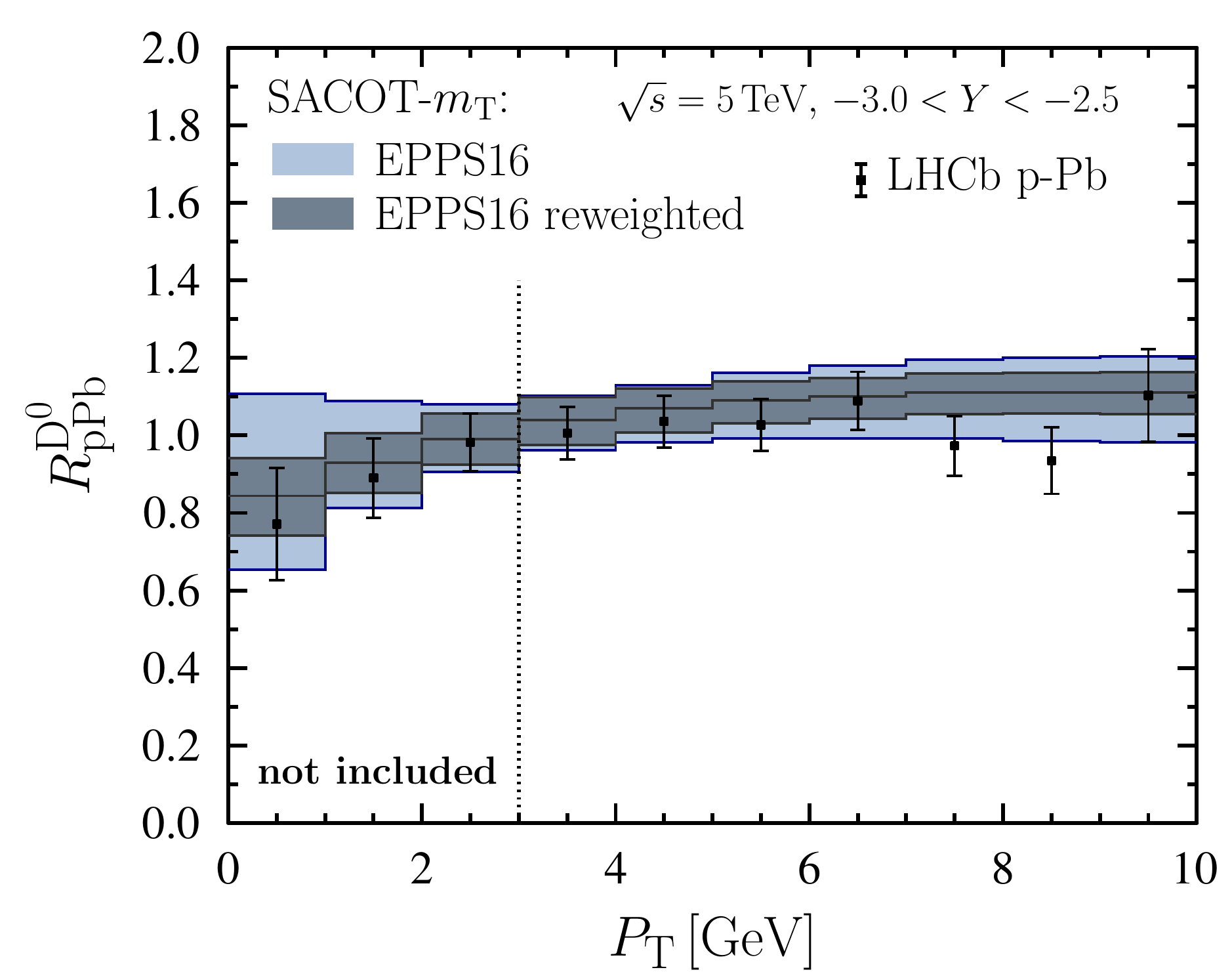}
\includegraphics[width=0.48\linewidth]{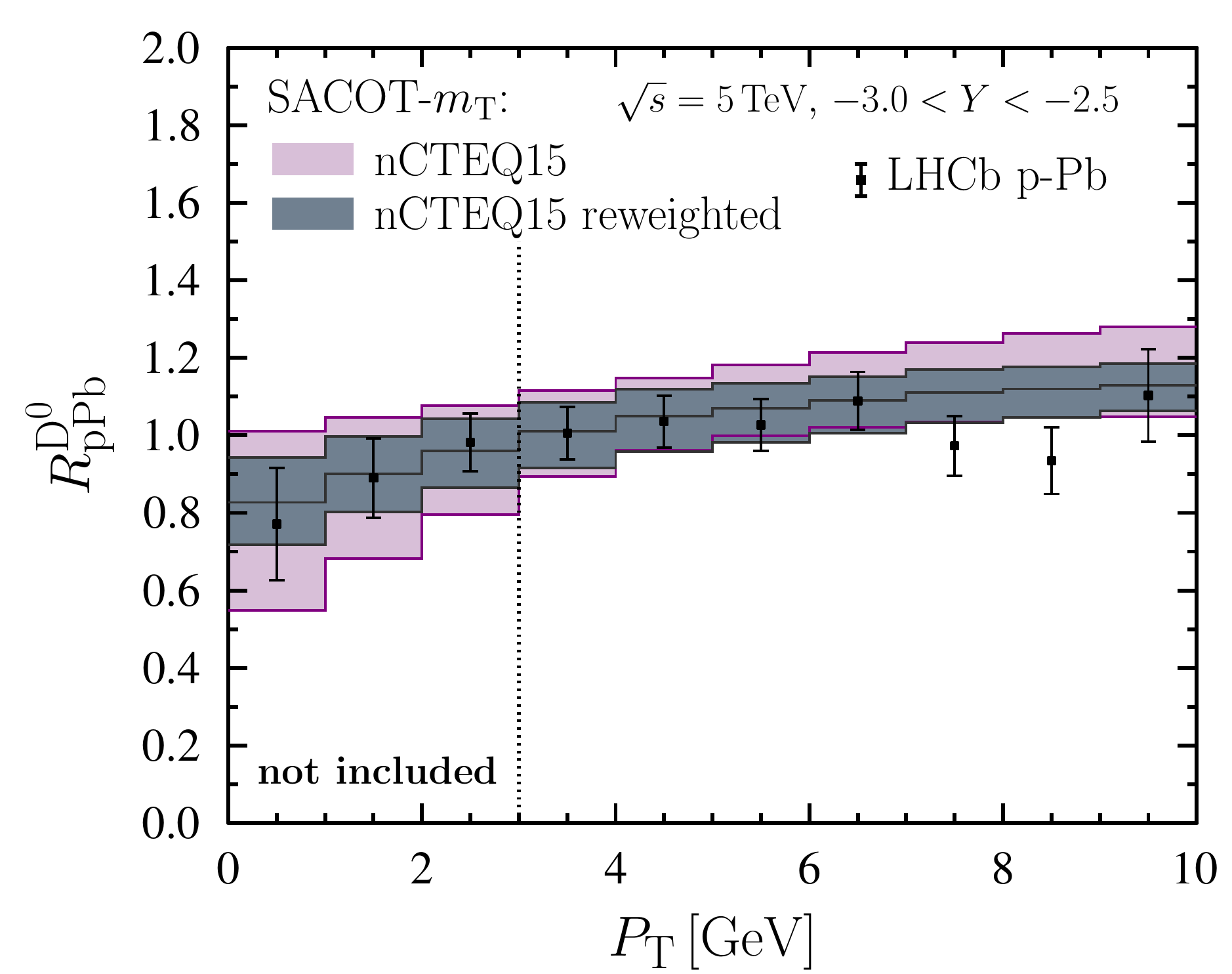} \\
\includegraphics[width=0.48\linewidth]{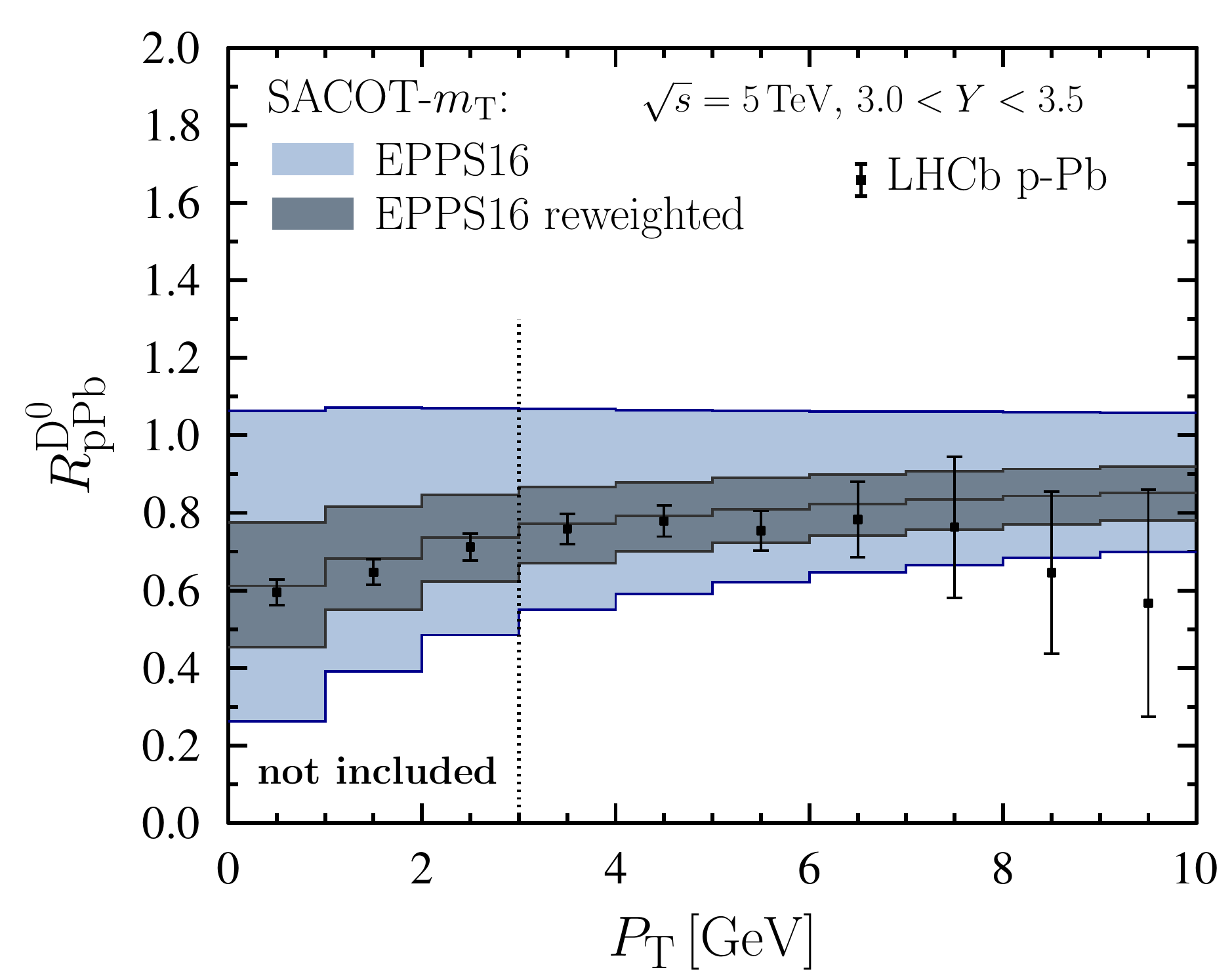}
\includegraphics[width=0.48\linewidth]{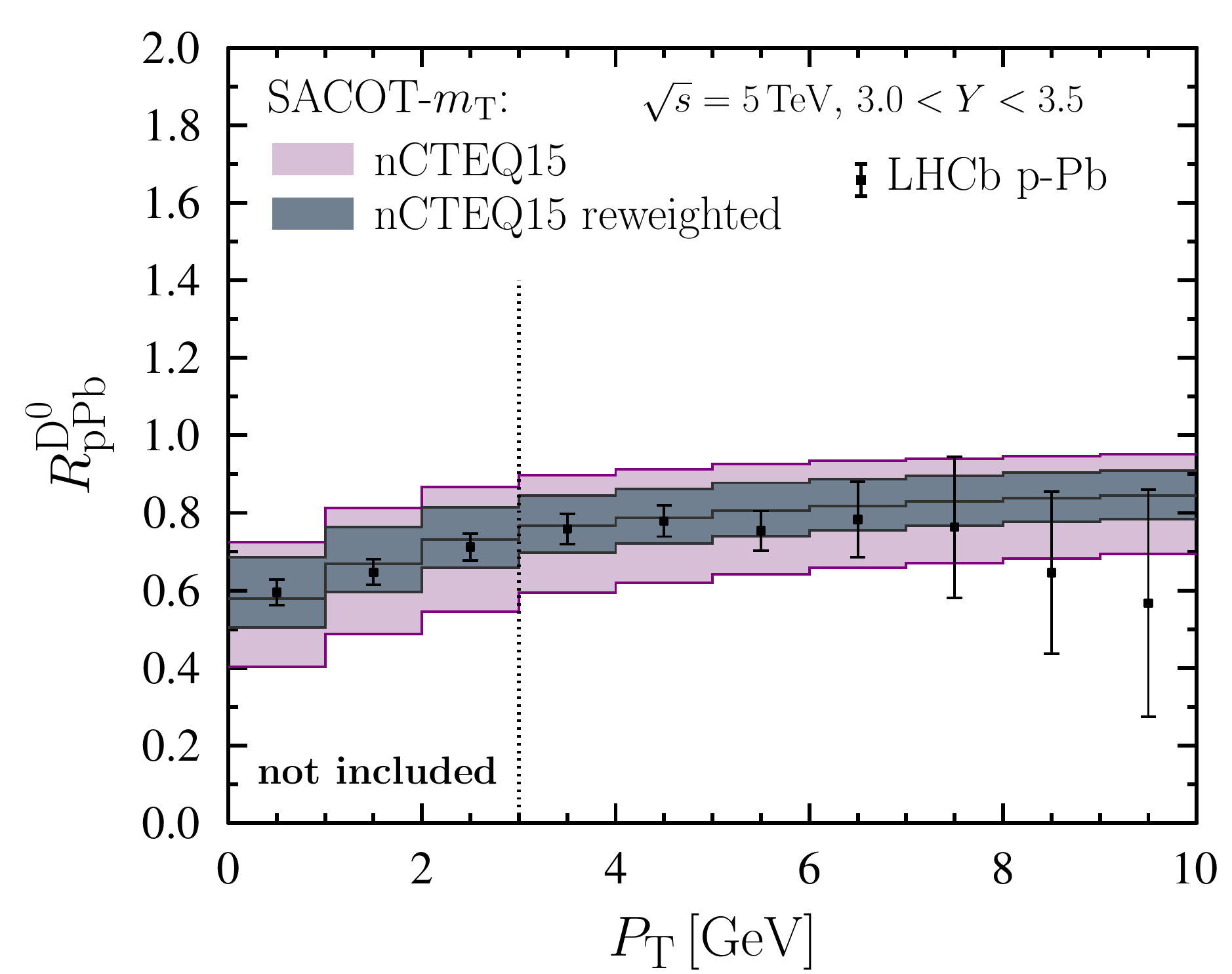}
\caption{As Figure~\ref{fig:pPb1} but showing also the results after reweighting.}
\label{fig:pPb2}
\end{figure}

\begin{figure}
\center
\includegraphics[width=0.45\linewidth]{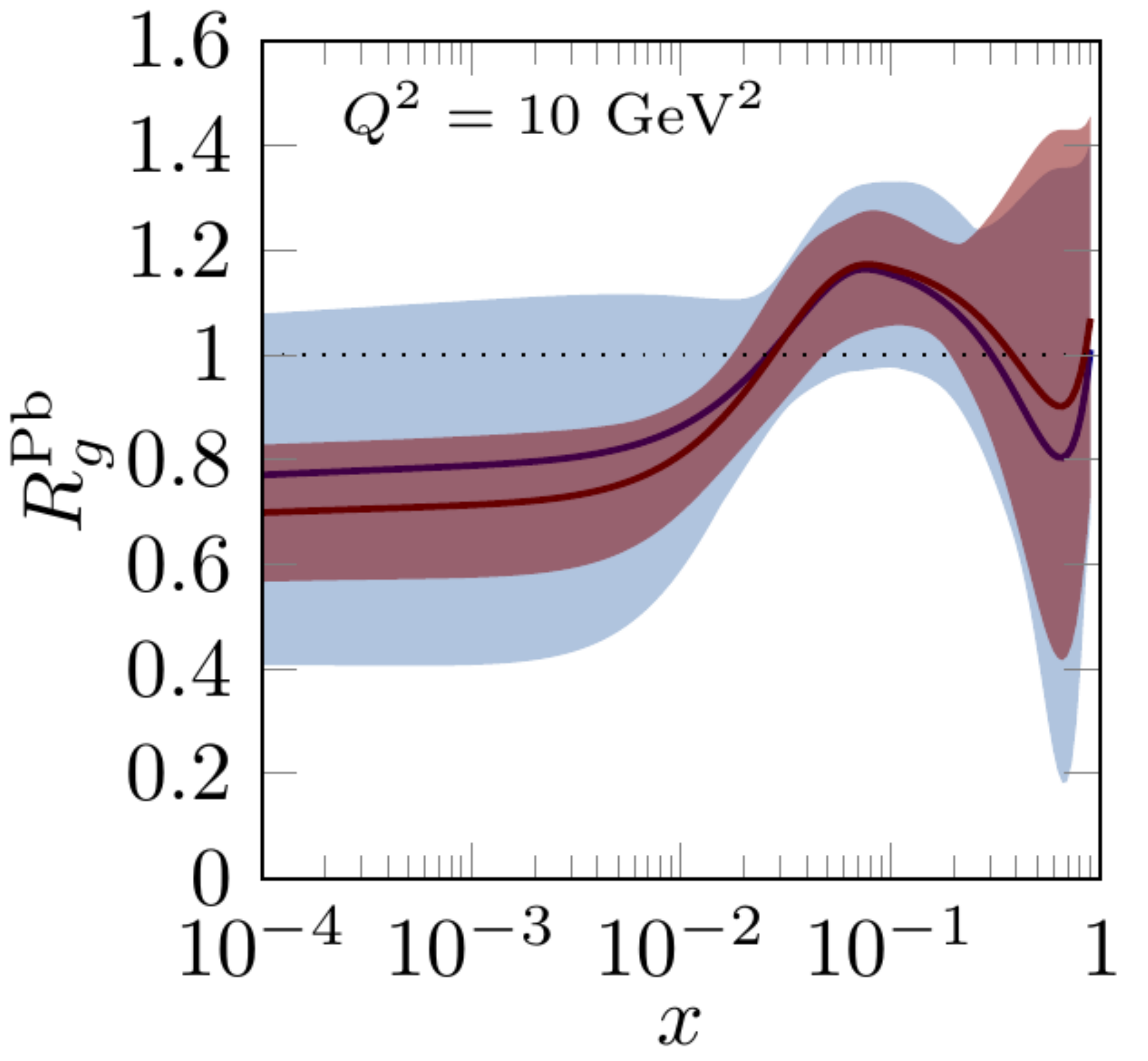}
\includegraphics[width=0.45\linewidth]{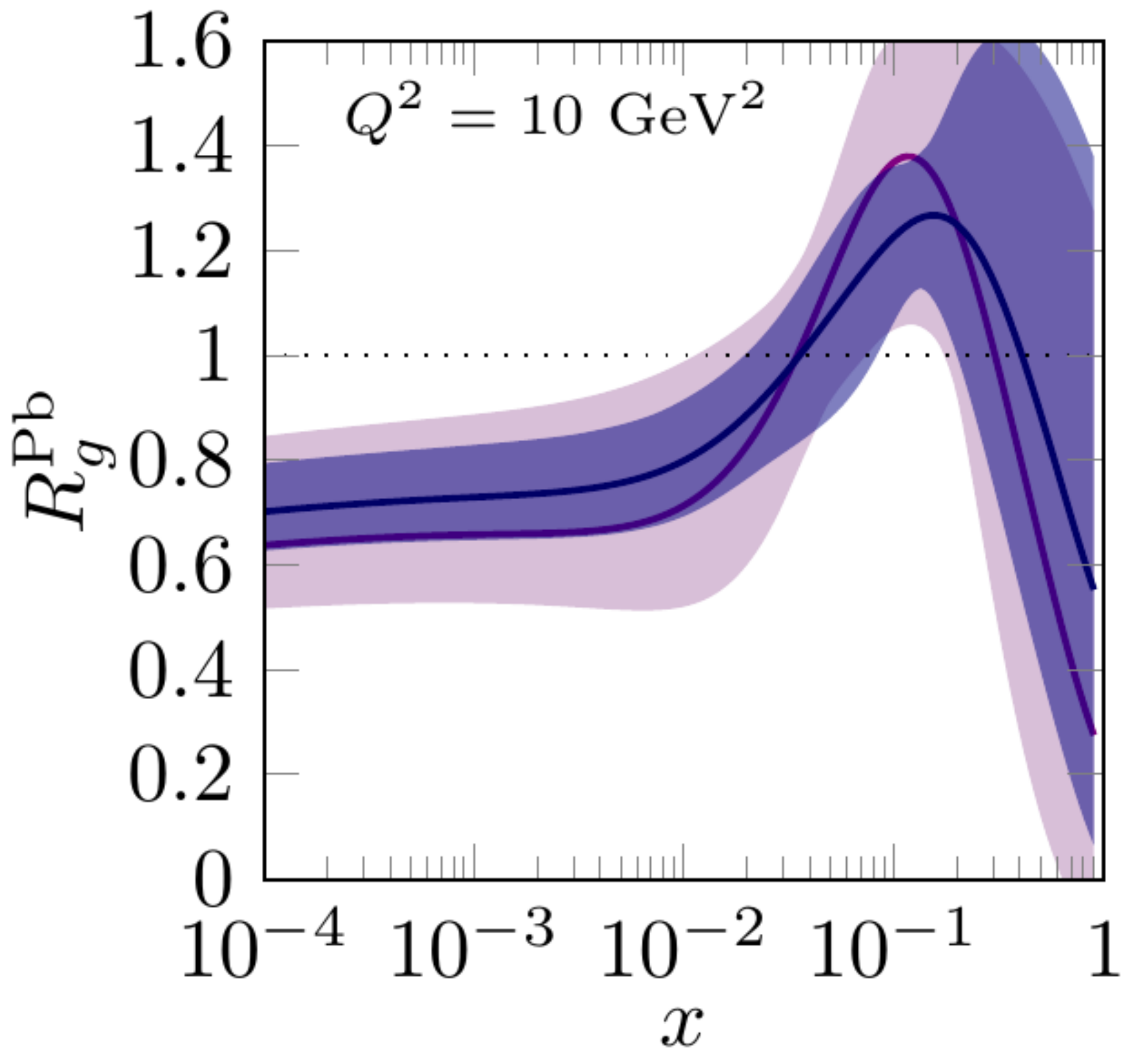} \\

\vspace{-2.2cm}
\hspace{0.8cm}\includegraphics[width=0.28\linewidth]{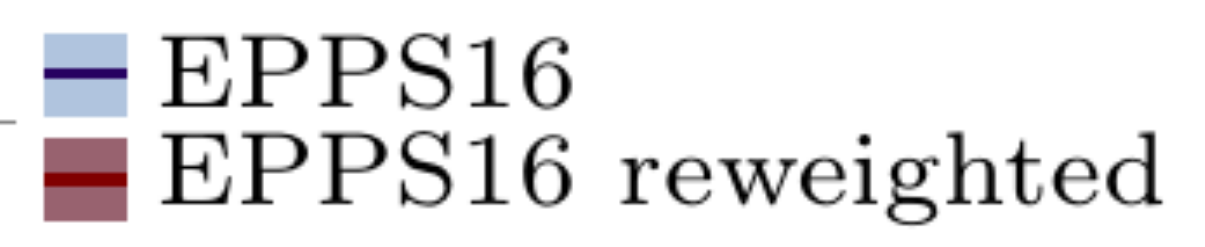}
\hspace{2.7cm}\includegraphics[width=0.28\linewidth]{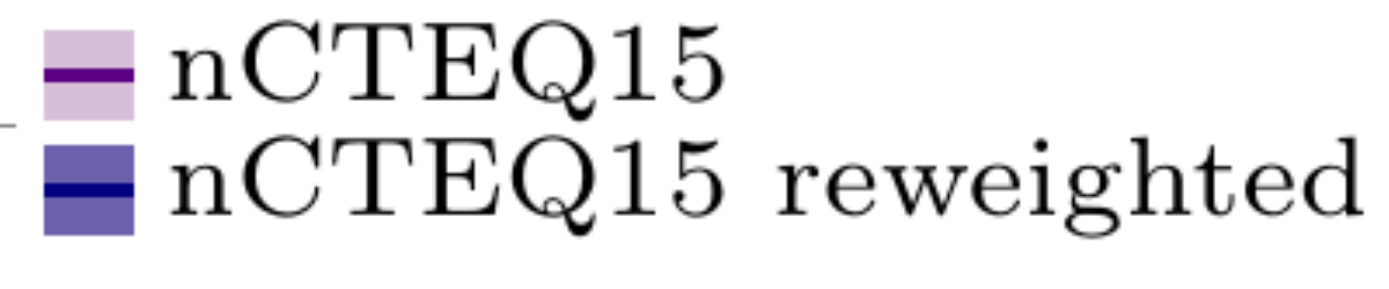}
\vspace{1.2cm}
\caption{The nuclear modifications as given by EPPS16 (left) and nCTEQ15 (right) parametrizations at scale $Q^2=10\,{\rm GeV}^2$ before and after reweighting.}
\label{fig:glues}
\end{figure}

Encouraged by the consistent description in Figure~\ref{fig:pPb1}, we have performed a Hessian reweighting analysis \cite{Paukkunen:2014zia,Eskola:2019dui} to check the impact these LHCb data would have on EPPS16 and nCTEQ15 \cite{Eskola:2019bgf}. Based on our observations in the preceding section, we have excluded the region $P_{\rm T} < 3\,{\rm GeV}$ from the analysis. Before the reweighting, we have $\chi^2/N_{\rm data} = 1.56$ for EPPS16 and $\chi^2/N_{\rm data} = 2.09$ for nCTEQ15, with $N_{\rm data}=48$ from altogether 8 different $Y$ bins. After the reweighting, these numbers improve to $\chi^2/N_{\rm data} = 1.02$ for EPPS16 and $\chi^2/N_{\rm data} = 1.12$ for nCTEQ15, so the description of the data gets better and both results are statistically reasonable. Figure~\ref{fig:pPb2} demonstrates the new PDF errorbands after the reweghting. The new errorbands are within the old ones which indicates that the reweighted PDFs are broadly consistent with the original ones. Especially in the case of EPPS16 with larger errors to begin with, the reduction in the PDF errors is quite dramatic. Even though we did not include the data points at $P_{\rm T} < 3\,{\rm GeV}$, we can see from Figure~\ref{fig:pPb2} that the data in this region are still very well described by both, reweighted EPPS16 and nCTEQ15. In other words, we do not find support for e.g. an onset of strong non-linear effects beyond the collinear factorization. Finally, Figure~\ref{fig:glues} shows the effect of reweighting on the gluon nuclear modifications for Pb. The reduction in the uncertainties is particularly notable for EPPS16 whose central value moves only very little upon reweighting. In the case of nCTEQ15 the reduction of the error band is also significant and we observe a preference for a milder excess (antishadowing) near $x=0.1$. All in all, we can conclude that the LHCb D$^0$ measurements can be included in the global fits of nuclear PDFs with no obvious clash with other data, and that they will have a significant impact especially on the gluon PDFs.

\vspace{-0.2cm}
\section*{Acknowledgments}
\vspace{-0.2cm}

We wish to acknowledge Academy of Finland Projects 297058 \& 308301, the Magnus Ehrnrooth Foundation, the Carl Zeiss Foundation, and the state of Baden-W\"urttemberg through bwHPC, for support. We have used the computer cluster of the Finnish IT Center for Science (CSC) in the presented work.

\end{document}